# Electrical Behavior of Downburst-Producing Convective Storms over the Western United States

Kenneth L. Pryor
Center for Satellite Applications and Research (NOAA/NESDIS)
Camp Springs, MD

**Abstract**

A great body of research literature pertaining to microburst generation in convective storms has focused on thermodynamic factors of the pre-convective environment as well as storm morphology as observed by radar imagery. Derived products based on GOES sounder data have been found to be especially useful in the study of thermodynamic environments. However, addressed much less frequently is the relationship between convective storm electrification, lightning phenomenology and downburst generation. Previous research in lightning production by convective storms has identified that electrification, phenomenology (i.e. flash rate, density), and polarity are dependent upon the thermodynamic structure of the ambient atmosphere, especially vertical moisture stratification. Thus, relevant parameters to describe the thermodynamic setting would include convective available potential energy (CAPE), due to its influence on updraft strength, and cloud liquid water content, due to its relationship to precipitation physical processes. It has already been addressed that buoyant energy and moisture stratification are important factors in convective storm development and downburst generation. This research effort investigates and derives a qualitative relationship between lightning phenomenology in convective storms and downburst generation. Downburst-producing convective storms will be analyzed by comparing pre-convective environments, as portrayed by GOES microburst products, storm morphology, as portrayed by radar imagery, and electrical behavior, as indicated by NLDN data.

## 1. Introduction

A great body of research literature pertaining to microburst generation in convective storms has focused on thermodynamic factors of the pre-convective environment as well as storm morphology as observed by radar imagery. Derived products based on GOES sounder and imager data have been found to be especially useful in the study of thermodynamic environments. However, addressed much less frequently is the relationship between convective storm electrification, lightning phenomenology and downburst generation. Previous research in lightning production by convective storms has identified that electrification, phenomenology (i.e. flash rate, density), and polarity are dependent upon the thermodynamic structure of the ambient atmosphere, especially vertical moisture stratification. Thus, relevant parameters to describe the thermodynamic setting would include convective available potential energy (CAPE), due to its influence on updraft strength, and cloud liquid water content, due to its relationship to precipitation physical processes. It has already been addressed that buoyant energy and moisture stratification are important factors in convective storm development and downburst generation.

The basic concept of charge generation and separation in convective clouds maintains that the presence of strong updrafts and the resulting development of precipitation are instrumental in the formation of an electric field of sufficient intensity for lightning discharge. In a convective cloud that builds to a height well above the freezing level, intense vertical motion will also loft large amounts of ice crystals to near the storm top. At the same time, precipitation, in the form of graupel or aggregates, will begin to descend within the convective cloud through the loading process. The interaction between ice crystals and larger precipitation particles (i.e. graupel, aggregates) will result in the acquisition of opposite electric charge between the lighter ice particles and the heavier precipitation particles, establishing a dipole and an electric field intensity necessary for electrical breakdown and subsequent lightning initiation (Saunders 1993). The descending precipitation within the storm is also vital for the development of convective downdrafts, accelerated as drier air from outside the convective storm cell is entrained. Thus, the physical process responsible for the initiation of lighting within a convective storm is also believed to be instrumental in the initiation of convective downdrafts that eventually produce downbursts at the surface.

Pryor (2006) developed a Hybrid Microburst Index (HMI) that infers the presence of a convective boundary layer (CBL) by incorporating the sub-cloud temperature lapse rate as well as the dew point depression difference between the typical level of a convective cloud base and the sub-cloud layer. Thus, the HMI algorithm describes the moisture stratification of the sub-cloud layer that may result in downdraft acceleration due to evaporational cooling of precipitation, eventually producing a downburst when the convective downdraft impinges on the earth's surface. In addition, Pryor (2008) recently developed a Geostationary Operational Environmental Satellite (GOES)-West (GOES-11) imager microburst algorithm that employs brightness temperature differences (BTD) between band 3 (upper level water vapor, 6.7μm), band 4 (longwave infrared window, 10.7μm), and split window band 5 (12μm). Band 3 is intended to indicate mid to upper-level moisture content and advection while band 5 indicates low-level moisture content. It follows that large BTDs between bands 3 and 5 imply a large relative humidity gradient between the mid-troposphere and the surface, a condition favorable for strong convective downdraft generation due to evaporational cooling of precipitation in the deep sub-cloud layer. In addition, small BTDs between bands 4 and 5 indicate a relatively dry surface layer with solar heating in progress. Large output BTDs in the GOES imager microburst product would most likely be associated large HMI values and an "inverted-v" sounding profile.

Accordingly, lightning data imagery, generated by a program that ingests the data from the National Lightning Detection Network (NLDN) to be plotted and mapped by Man computer Interactive Data Access System (McIDAS), has been collected for several severe convective storm events that occurred in the inter-mountain western United States during the summer of 2005. For each downburst occurrence, NLDN data has been compared to radar and satellite imagery and surface observations of downburst winds. It has been observed that suppressed cloud-to-ground (CG) lightning flash rates were associated with downburst occurrence within a high-reflectivity convective storm that was documented over southeastern Idaho in August 2005. The environment was characterized by the presence of a relatively deep and dry CBL, as indicated by GOES sounding profile data, the GOES imager microburst product, and surface observations. Also observed with this storm was the dominance of positive cloud-to-ground lightning discharges (+CGs), especially near the time and location of downburst occurrence. The reader is referred to MacGorman and Rust (1998) and Uman (2001) for a discussion of convective storm electrification, and lightning categorization and phenomenology.

It has been suggested that decreased CG flash rates in convective storms may be a consequence of an elevated charge dipole, driven by especially vigorous updrafts, that results in reduced electric field intensity between the convective cloud and the surface. It follows that more intense updrafts would increase precipitation content within the convective cell, thereby enhancing the effect of precipitation loading (Doswell 2001; Wakimoto 2001). Once the process of precipitation loading has initiated a downdraft, entrainment of dry (or low theta-e) air in the midlevels of the storm cell would enhance downdraft strength by the process of evaporative cooling. Thus, periods of enhanced updraft intensity within a convective storm could be associated with an increased likelihood for downburst generation, and, hence, imply a relationship between CG flash rates and downburst occurrence.

This research effort investigates and derives a qualitative relationship between lightning phenomenology in convective storms and downburst generation. Downburst-producing convective storm systems will be analyzed by comparing pre-convective environments, as portrayed by GOES sounding profiles and the imager microburst products; storm morphology, as portrayed by radar imagery; and electrical behavior, as indicated by NLDN data. In addition, this paper has been revised to include relevant imagery from the recently developed GOES imager microburst product. For the 10 August 2005 convective storm event over southeastern Idaho, GOES imager microburst products will be compared to coincident sounding profile data to identify and illustrate the environmental profile favorable for low flash-rate, positive stroke dominated (PSD) storms and resultant downburst generation.

## 2. Methodology

A convective storm event over southeastern Idaho was selected for study to accomplish a twofold objective: to investigate the electrical behavior of downburst-producing convective storms over the inter-mountain western United States as well as to validate the new GOES imager microburst product. The reader is referred to Pryor (2006) and Pryor (2008) for a description of the HMI algorithm and the GOES imager microburst product, respectively. National Lightning Detection Network (NLDN) data was collected and then plotted and mapped by the Man computer Interactive Data Access System (McIDAS) for a convective storm that developed and evolved over southeastern Idaho on 10 August 2005. For this downburst event, NLDN data was compared to radar and satellite imagery and surface observations of downburst winds. The number of cloud-to-ground (CG) lightning discharges was determined for each storm. Flashing rates, averaged over 30-minute intervals, were then calculated based on each flash count. Microburst product data were validated against conventional surface data as recorded by

Idaho National Laboratory (INL) mesonet stations in southeastern Idaho. The 33-station mesoscale surface observation network at INL was selected in this study by virtue of its high temporal and spatial resolution as well as the high quality of meteorological observations that encompass the Snake River Plain of Southeastern Idaho. Clawson et al. (1989) provides a detailed climatology and physiographic description of the INL as well as a description of the associated meteorological observation network.

HMI product imagery was not available for this event. In lieu of HMI data, a representative HMI value was manually calculated based on the GOES sounding profile generated at Idaho Falls, Idaho at 2300 UTC 10 August 2005. Product images were generated in Man computer Interactive Data Access System (McIDAS) by a program that reads and processes GOES imager data, calculates brightness temperature differences, and overlays risk values on GOES imagery. The image data consisted of derived brightness temperatures from infrared bands 3, 4, and 5, obtained from the Comprehensive Large Array-data Stewardship System (CLASS, http://www.class.ncdc.noaa.gov/).  In order to assess the predictive value of GOES microburst products, GOES data used in validation was obtained for the nearest retrieval time to the observed surface wind gusts.

Downburst wind gusts, as recorded by National Oceanic and Atmospheric Administration (NOAA) INL mesonet stations, were measured at a height of 15 meters (50 feet) above ground level. Archived NOAA mesonet observations are available via the NOAA INL Weather Center website (http://niwc.noaa.inel.gov).  Next Generation Radar (NEXRAD) base reflectivity imagery (levels II and III) from National Climatic Data Center (NCDC) was utilized to verify that observed wind gusts were associated with high-reflectivity downbursts and not associated with other types of convective wind phenomena (i.e. gust fronts). NEXRAD images were generated by the NCDC Java NEXRAD Viewer (Available online at http://www.ncdc.noaa.gov/oa/radar/jnx/index.html). Another application of the NEXRAD imagery was to infer microscale physical properties of downburst-producing convective storms. Particular radar reflectivity signatures, such as the rear-inflow notch (RIN)(Przybylinski 1995) and the spearhead echo (Fujita and Byers 1977), were effective indicators of the occurrence of downbursts.  In addition, echo tops (ET) data were collected and analyzed to infer convective storm physical characteristics, serving as a proxy for updraft intensity. The selection of this radar data for analysis was based on the premise that higher echo tops are associated with more intense convective updrafts.

## 3.  Case Study:  10 August 2005 Idaho Downbursts

During the afternoon of 10 August 2005, a multi-cellular convective storm initiated over the White Knob Mountains of south-central Idaho, west of the Idaho National Laboratory (INL) mesonet. Orographic lift served as an initiating mechanism for the convective storm complex. Between 1900 and 2000 UTC, radar reflectivity imagery (not shown) displayed the development and evolution of the storm complex.

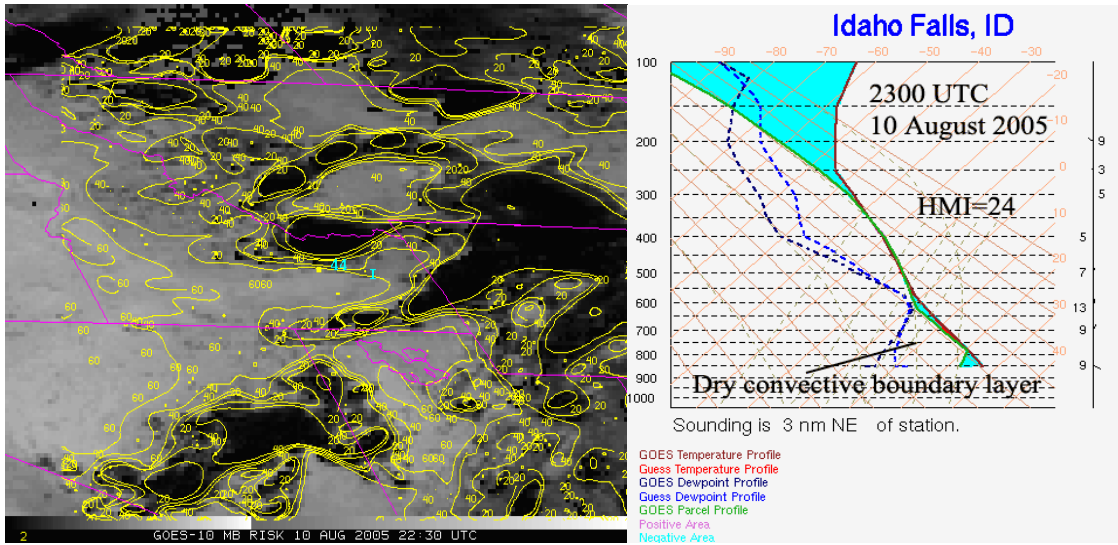

Figure 1. GOES-11 imager microburst risk product at 2230 UTC 10 August 2005 (top) and GOES sounding profile at Idaho Falls, Idaho at 2300 UTC 10 August 2005 (bottom). Location of GOES sounding is marked "I" in the microburst risk image.

Output BTDs between 40°C and 50°C in the vicinity of Rover mesonet station at 2230 UTC are displayed in Figure 1. In addition, a large surface dewpoint depression of 23°C, recorded by the Rover observing station at 2245 UTC, just prior to downburst occurrence and shown in Figure 2, indicated that a convective boundary layer (CBL) was well established over the Upper Snake River Plain. Stull (1988) noted that a large surface dewpoint depression (> 17°C) is associated with a well-developed CBL. Also shown in Figure 1, the 2300 UTC 10 August 2005 GOES sounding profile from Idaho Falls, is minimal CAPE and the presence of a deep, well-mixed boundary layer and very dry surface layer. The "inverted-v" sounding profile as well as the calculated HMI value of 24 based on the sounding indicates that sub-cloud evaporation of precipitation was an important factor in convective downdraft acceleration. Comparing output BTDs at Idaho Falls in the microburst risk image in Figure 1 to the sounding profile exemplifies the close correspondence between high BTDs and the "inverted-v" profile. Also important to note is the level of the -20°C isotherm at the height of the 400-mb level (19586 feet (~6000 m)).

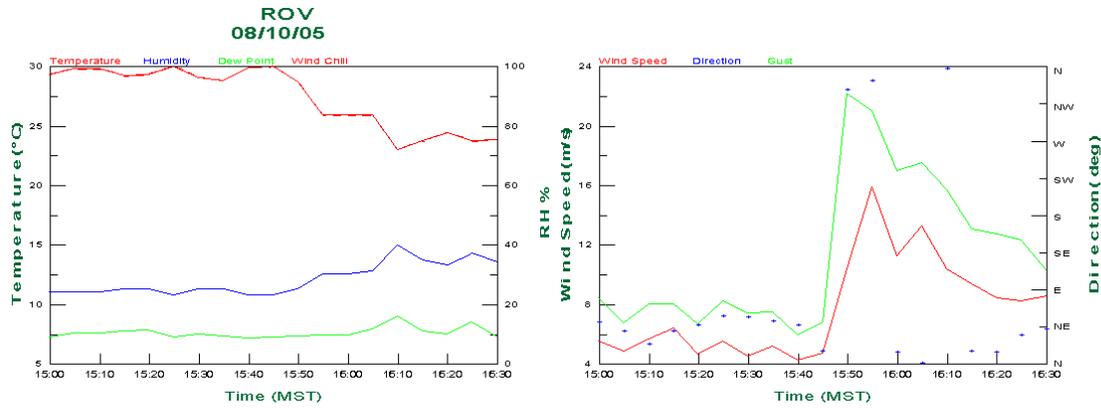

Figure 2. INL mesonet meteogram from Rover observing station displaying temperature, humidity, and wind conditions between 2200 and 2330 UTC (1500 and 1630 MST) 10 August 2005.

Radar imagery from the Pocatello, Idaho WSR-88D displayed the evolution of the southernmost cell in the convective storm cluster into a spearhead echo as the cell moved from Butte County eastward into Jefferson County within the INL between 2231 and 2246 UTC (not shown). The convective storm rapidly intensified as reflectivities increased to over 65 dBZ by 2251 UTC, as portrayed in Figure 3. In addition to increasing reflectivity, rapid increases in echo tops (to over 40000 feet (~12000 m)) signified a significant increase in updraft intensity and precipitation content of the convective storm, precursors to downburst generation.

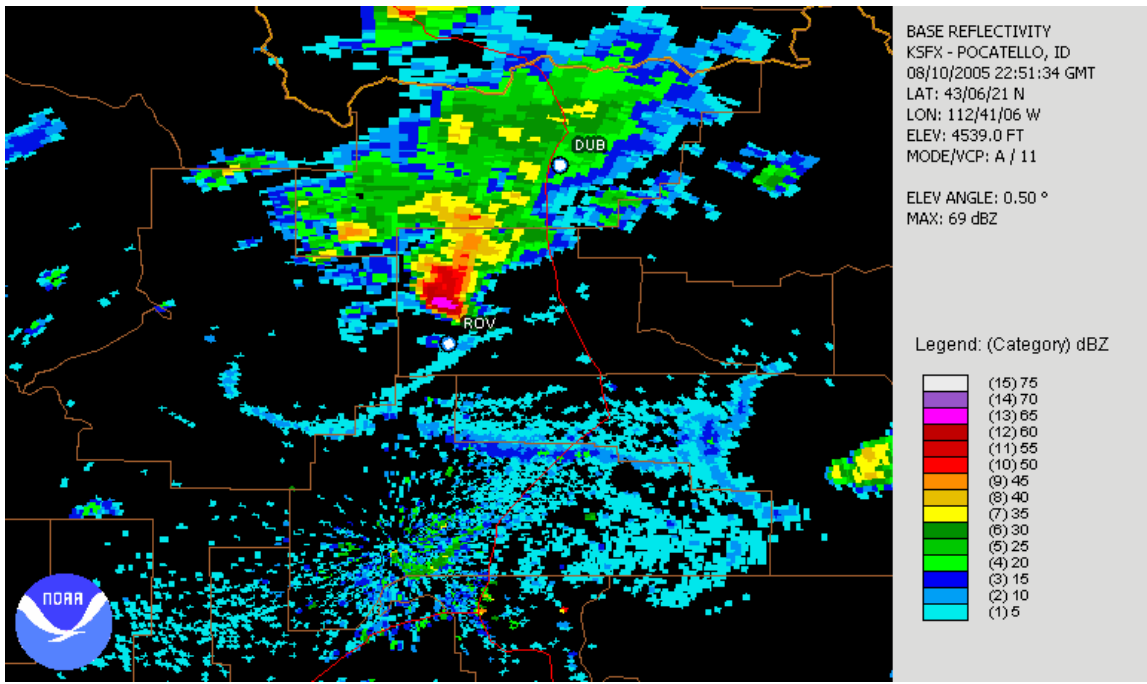

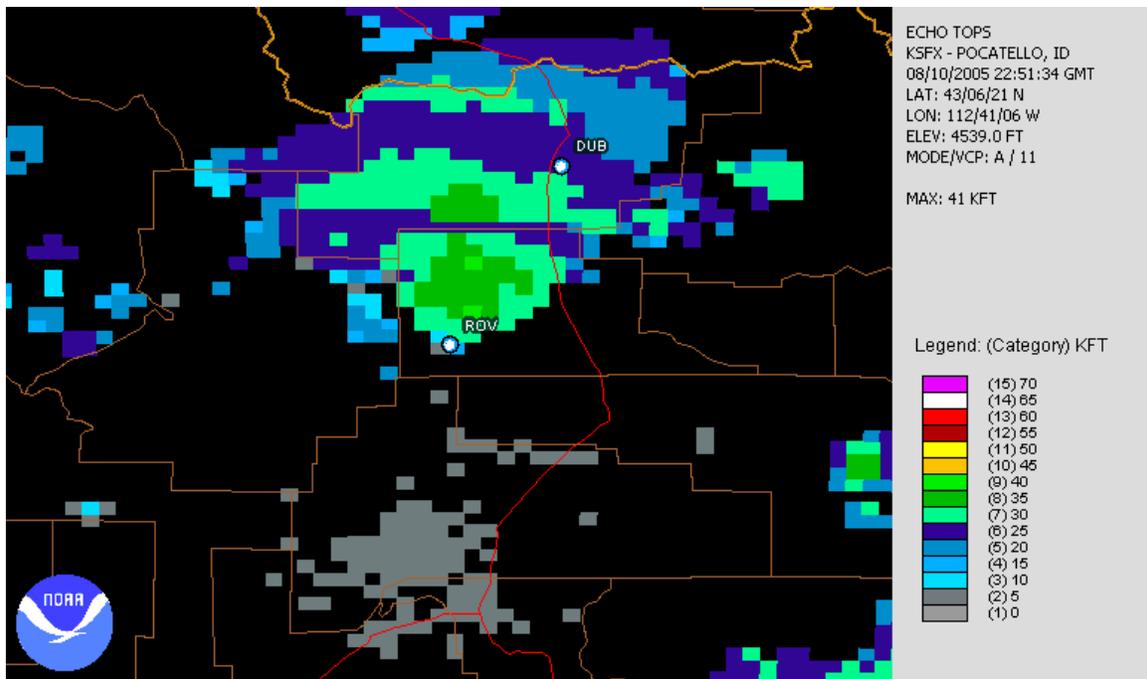

Figure 3. Pocatello, Idaho NEXRAD reflectivity imagery (top) and echo tops (bottom) at 2251 UTC 10 August 2005. Blue-outlined marker indicates location of Rover observing station.

Examination of radar imagery at various elevation angles (not shown) revealed reflectivities > or = 30 dBZ extending to a height of approximately 35000 feet (~10668 m). Another indicator of the presence of strong convective updrafts is the displacement of maximum echo tops to the eastern periphery of the storm, over the low-level reflectivity gradient and away from the location of maximum radar reflectivity (Przybylinski and

Gery 1983). At 2250 UTC, a downburst wind gust of 44 knots was recorded by the Rover observing station. Apparent in radar imagery at 2251 UTC (Figure 3), near the time of microburst occurrence, was the evolution of the southernmost cell into a spearhead echo configuration.

Fujita and Byers (1977) described the spearhead echo as "a radar echo with a pointed appendage extending toward the direction of the echo motion. The appendage moves faster than the parent echo which is being drawn into the appendage. During the mature stage, the appendage turns into a major echo and the parent echo loses its identity". Fujita and Byers (1977) associated the spearhead echo with downburst generation in particularly intense convective storms. High radar reflectivities occurring in an environment characterized by a deep, dry convective mixed layer suggests that this microburst can be classified as a hybrid microburst.

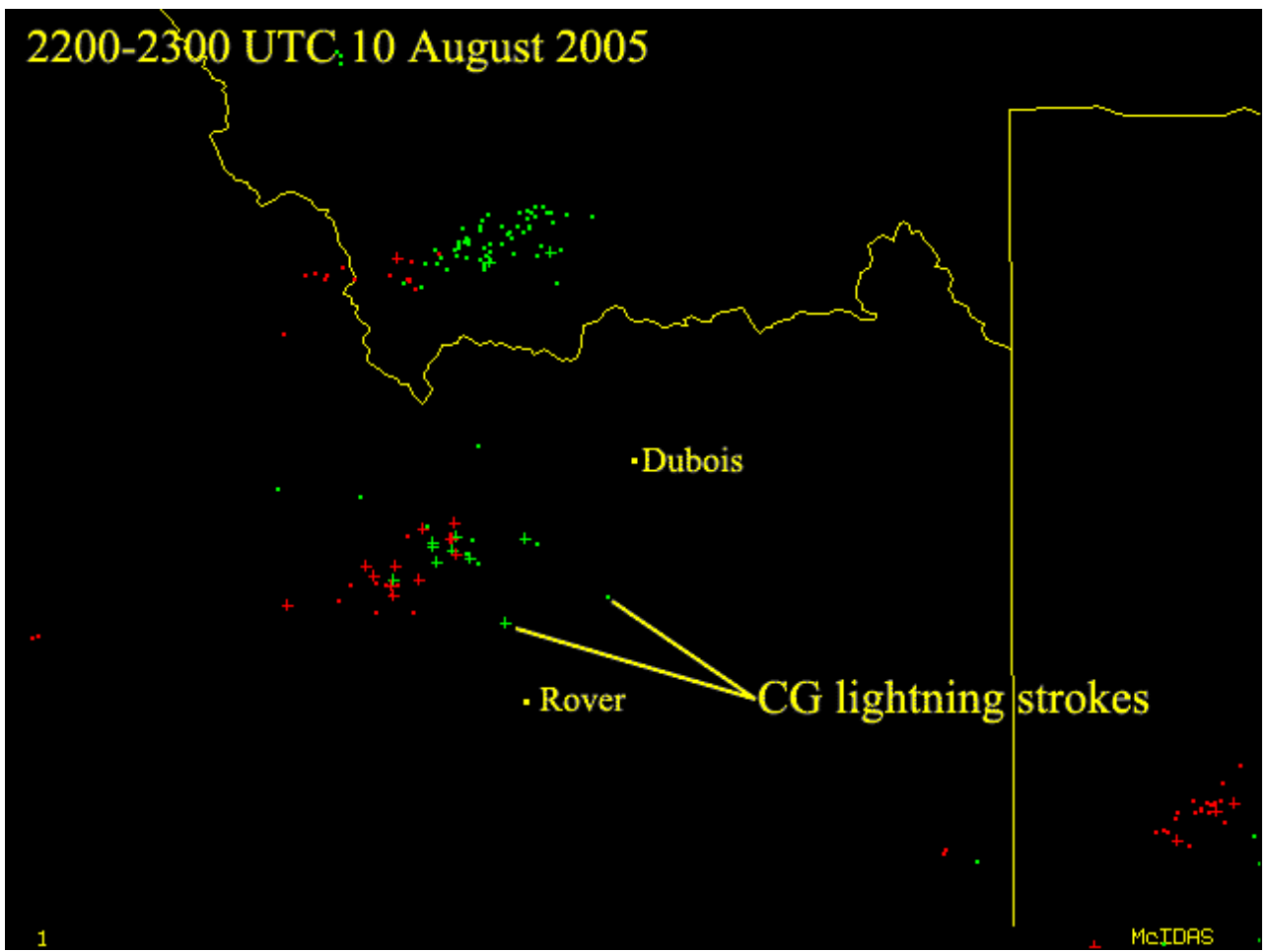

Figure 4. National Lightning Detection Network (NLDN) image displaying cloud-to-ground (CG) discharges between 2200 and 2300 UTC 10 August 2005. Data plotted in color red indicates discharges that occurred during the period 2200 to 2230 UTC while data plotted in color green indicates discharges that occurred during the period 2230 to 2300 UTC. A yellow marker indicates the location of Rover observing station.

Also interesting, as shown in Figure 4, during the 30-minute interval encompassing the microbursts (2230 to 2300 UTC), a decreased CG flash rate (.6/min) was recorded by the NLDN in association with the convective storm cluster. Also noteworthy is a slight decrease in echo tops after microburst occurrence. During the following 30-minute interval (not shown), 2300 to 2330 UTC, the storm CG flash rate increased to 1.1/min in conjunction with intensification of a cell in the eastern portion of the storm complex. However, no further microburst activity occurred in association with the convective storm. Overall, the storm was characterized with a low flash rate, increasing from .63/min to 1.1/min between the time intervals 2230 to 2300 and 2300 to 2330 UTC. During the one hour period 2200 to 2300 UTC, the convective storm was positive stroke dominated (PSD) with 53 percent of the flashes recorded as positive by the NLDN.

The sign of CG lightning discharge (positive or negative) and the CG flash rate could be useful in this case to infer attributes of the ambient environment of the convective storm. The very low CG flash rate during the time of downburst generation most likely indicates the presence of an elevated dipole. The very dry boundary layer as well as the presence of dry air in the mid-troposphere is believed to have influenced the nature of the electrification of the convective storm cell. For example, it has been noted in previous research that PSD storms are the result of an inverted dipole (negative charge center over positive charge center) (Williams 2001). Dry (low theta-e) air in the mid-troposphere as well as in a deep layer below the cloud base may be responsible for the development of this inverted dipole. When entrained into the convective storm cell, the drier air will cause the opposite charge structure to a normal polarity storm: graupel or aggregates, in the storm mid-levels charge positive, while ice crystals, lofted to the upper levels of the storm, charge negative (Saunders 1993). The 2300 UTC 10 August 2005 GOES sounding from Idaho Falls provides evidence of this favorable moisture stratification for PSD storm development with a dry air layer in the mid-troposphere, around the 400-mb level, and the presence of a deep, dry convective boundary layer. Further supporting evidence of the development of an elevated, inverted dipole could be the radar observation of the precipitation core (reflectivities > or = 30 dBZ) extending above the level of the -20°C isotherm as indicated by the sounding profile. During the 30-minute period after the termination of microbursts, there was a considerable increase in CG flash rate. It is speculated that this increase in flash rate was attributable to weakening updrafts and the subsequent descent of the positive charge center, increasing the electric field intensity between the cloud and the ground. The electrical behavior of this storm is typical of microburst-producing convective storms as discussed by Williams (2001).

This case demonstrated the influence of static instability and moisture stratification on both lightning and downburst generation. Again, the reduced flash rate associated with microburst activity was most likely the result of an elevated dipole. McCaul and Cohen (2002) have noted that in CAPE-starved environments, a high lifting condensation level (LCL) that is roughly equal to a corresponding level of free convection (LFC) favor the generation of strong updrafts due to compression of buoyancy. The authors also noted that an optimal LFC for updraft overturning efficiency is greater than 2 km (6561 feet) AGL. This case exemplifies their finding, with an LFC and LCL near the 600 mb level at approximately 9000 feet AGL.  Thus, the elevated dipole hypothesized in this case was likely the result of buoyancy compression and the subsequent generation of strong convective updrafts.  In addition, as discussed in Saunders (1993), moisture stratification and convective instability favored low flash rate, PSD storms as explained by the following tentative factors:

1. Charging of precipitation particles, in the form of graupel (or aggregates) and ice crystals, and the attendant separation interactions most likely took place at a higher level within the storm where temperatures were at or below -20°C. Thus, graupel and/or aggregates charged positive while ice crystals charged negative. Radar indicated echo tops (over 40000 feet (12000 m)) well above the level of the -20°C isotherm supports this hypothesis.
2. Relatively low liquid water content as inferred by dryness in the boundary layer (surface dewpoints below 50°F (10°C)).
3. The presence of smaller precipitation particles (especially graupel and/or aggregates) that carried lower amounts of electric charge and the separation thereof, generating weaker electric field intensity, especially between the lower (positive) charge center within the cloud and the earth's surface.

## 4.  Summary and Conclusions

Based on analysis of pre-convective environments using GOES sounder data and derived products, and review of literature pertaining to convective storm electrification, proposed physical explanations are offered that describe a qualitative relationship between lightning and downburst generation. The Idaho downburst storm developed and evolved in an environment characterized by a dry convective boundary layer with surface dewpoints below 50°F (10°C). An elevated, inverted charge dipole is believed to have effected a low lightning flash rate that accompanied downburst occurrence in this convective storm. Radar indicated echo tops that exceeded 40000 feet (12000 m) supported the inference of strong updrafts that may have elevated the level of precipitation development and the resulting charge separation.  It was detailed by Saunders (1993), that the positive charging of graupel is likely to occur with colder temperatures (at or below -20°C) that are typically found at higher levels within a convective storm cell. Thus, with this western U.S. convective storm, the hypothesized elevated charge dipoles that developed produced a decrease in the cloud-to-surface electric fields, thereby reducing the occurrence of CG lightning discharges, especially in the vicinity of downburst generation. Also, the resulting inverted dipole likely resulted in the dominance of positive CG strokes associated with downburst occurrence. Accordingly, this study has investigated and highlighted the effective use of various remote sensing resources, including GOES sounder and imager-derived products, radar imagery and NLDN data to assess downburst potential and occurrence with convective storms. It is apparent that, over the intermountain western United States, reduced lightning discharge rates accompany downburst generation. Therefore, the effective use of NLDN data in conjunction with conventional radar imagery can alert the operational forecaster to impending downburst occurrence.